# Dust-acoustic solitary waves in dusty plasma with variable dust charge

**Gh. Forozani[1] and M. Mohammadi**

Department of Physics, Bu_Ali Sina University, Hamedan-Iran

**Abstract**

In this article we are going to consider dust acoustic wave in dusty plasma whose constituents are inertial negative charged dust particles ,Boltzmann distributed electrons and non-thermal distributed ions with variable dust charge. Using reductive perturbation method, we have obtained Korteweg-de Veries (kdv) and modified kdv(mkdv) equations. A Sagdeev potential for the system and stability conditions for solitonic solution are also derived.
PACS 50

**1. Introduction**

The creation of dusty plasma is because of entry dust particles with nano up to one micrometer size that have large masses which supports a variety of collective phenomena that have been the subject of a large number of investigations recently [1]. The charged dust grains bring about many significant changes in behavior of the system including the creation of new modes. The dust acoustic wave(DAW) is such one new mode[2]. Dust acoustic waves were first reported theoretically in unmagnetized dusty plasma by Rao et. al [3] . On the other hand , at higher frequency, Shukla and Silin showed the existence of dust ion acoustic waves[4].

The dusty plasma is one of the most rapidly growing branches of plasma physics .In recent years most published papers in dusty plasma physics are related to dust acoustic waves and dusty plasma lattice. By using the reductive perturbation we obtained the propagation of dusty acoustic wave. In more investigations Korteweg-de Vries(kdv) , Zakharov-Kuznetsov (zk) and Kadomster – Petriashvili (kp) equations are investigated. The investigation result of propagation dust acoustic wave in one dimension[5-7] , is kdv equation and by consideration waves in higher dimensions kp equation is obtained[8]. The kdv equation has both mathematical and physical aspect. The kdv equation can be solved exactly through analytical methods or by using numerical method.

Lin and Daun also considered dust acoustic solitary wave in a dusty plasma with non-thermal ions[9] .In this paper we considered unmagnetized dusty plasma which including variation dust charge , Boltzmann distributed electrons and non –thermal distributed ions. In section 2 we will show the basic equation of dust acoustic wave motion and in section 3 by using reductive perturbation method we obtained kdv equation, as well as Sagdeev potential and stability condition for solitonic solution is investigated. In section 4 mkdv equation is obtained.

**2.The model equation**

---
[1] Corresponding author
E-mail addresses: forozani@basu.ac.ir.
Fax: +98 811 8280440
Tel: +98 811 8280440

Basic equation for consideration dust acoustic wave in unmagnetized dusty plasma with include variable dust charge , Boltzmann electrons and non-thermal ions , is given by:

$$\frac{\partial n_d}{\partial t} + \frac{\partial (n_d u_d)}{\partial x} = 0 \tag{1}$$

$$n_d \frac{\partial u_d}{\partial t} + n_d u_d \frac{\partial u_d}{\partial x} + \sigma_d \frac{\partial p_d}{\partial x} = -n_d z_d \frac{\partial \emptyset}{\partial x} \tag{2}$$

$$\frac{\partial p_d}{\partial t} + u_d \frac{\partial p_d}{\partial x} + 3p_d \frac{\partial u_d}{\partial x} = 0 \tag{3}$$

$$\frac{\partial^2 \emptyset}{\partial x^2} = z_d n_d + \mu n_e - \frac{1}{1-\mu} n_i \tag{4}$$

where $n_d$ Is the dust particle number density normalized to $n_{d0}$. $u_d$ is dust particle velocity normalized to $c_d = (\frac{z_d T_i}{m_d})^{\frac{1}{2}}$. $z_d$ , $m_d$ and $\emptyset$ number of charge on dust particles, particle mass and electrostatic potential, respectively. $z_d$ and $\emptyset$ are normalized to $z_{do}$ and $\frac{T_i}{e}$, respectively. Time and space variable are normalized respectively to the dust plasma period $w_{pd}^{-1} = (\frac{m_d}{4\pi e^2 n_{d0} z_d^2})^{\frac{1}{2}}$ and the Debye length $\lambda_{De} = (\frac{T_i}{4\pi e^2 n_{d0} z_d^2})^{\frac{1}{2}}$. $\sigma_i = \frac{T_i}{T_e}$ where $T_i$ and $T_e$ are temperature ions and electrons respectively and $\mu = \frac{n_{0e}}{n_{0i}}$, where $n_{0e}$ and $n_{0i}$ are number density of unperturbed ions and electrons in dust acoustic.

$n_i$ is number density of ions and $n_e$ is number density of electrons which normalized to $n_{0i}$ and $n_{0e}$ respectively. $n_i$ and $n_e$ is given by[10,11]:

$$n_i = \left[1 + \frac{4\alpha}{1+3\alpha}(\emptyset + \emptyset^2)\right] \exp(\emptyset) \tag{5}$$

$$n_e = \frac{1}{1-\mu} \exp(\sigma_i \emptyset) \tag{6}$$

$\alpha$ being a parameter defining the population of non –thermal ions [11,12] . . In the case of $\alpha = 0$ , we can neglect the effect of non-thermal ions .Total charge neutrality at equilibrium is:

$$n_{0e} + n_{0d} z_{0d} = n_{0i} \tag{7}$$

Where $n_{0e}$ , $n_{0i}$ , $n_{0d}$ and $z_{0d}$ are number density of unperturbed electrons, ions and the number density equilibrium values of unperturbed number of charge on the dust particles respectively. The Variation of dust charge is because of collision between electrons and ions with dust particles. The collision of electrons and ions with plasma particles produces a charged current. The charge-current balance equation is given by :

$$\frac{\partial q_d}{\partial t} = I_e + I_i \tag{8}$$

Where $I_e$ and $I_i$ are current of electrons and ions respectively. If we consider that the streaming velocities of electrons and ions are much smaller than the thermal velocities. Therefore $\frac{dq_d}{dt} \ll I_i, I_e$ and charge- current balance equation (8)reduce to equation $I_i + I_e \approx 0$.

The current of ions and electrons are[13]:

$$I_e = -e\pi r^2 \left(\frac{8T_e}{\pi m_e}\right)^{\frac{1}{2}} n_e \exp\left(\frac{e\Phi}{T_e}\right) \tag{9}$$

$$I_i = e\pi r^2 \left(\frac{8T_i}{\pi m_i}\right)^{\frac{1}{2}} n_i \left(1 - \frac{e\Phi}{T_i}\right) \tag{10}$$

where $\Phi$ denotes the dust particle surface potential related to the plasma potential $\emptyset$[14] and $Z_d$ is the normalized dust charge obtained from:

$$Z_d = \frac{\Psi}{\Psi_0} \tag{11}$$

Where $\Psi = \frac{e\Phi}{T_{eff}}$ and $\Psi_0 = \Psi(\emptyset = 0)$

### 3. Derivation of kdv equation

The propagation equation of the dust acoustic wave considering unmagnetized dust plasma containing the variety of dust particles may be investigated by introducing new coordinates $\xi$ and $\tau$ defined as:

$$\xi = \varepsilon^{\frac{1}{2}}(x - \lambda t) \, , \, \tau = \varepsilon^{\frac{3}{2}} t \tag{12}$$

where $\lambda$ Is the phase velocity and $\varepsilon$ is a small dimensionless expansion parameter. By expanding the independent variables we have:

$$n_d = 1 + \varepsilon n_{d1} + \varepsilon^2 n_{d2} + \varepsilon^3 n_{d3} + \cdots \tag{13}$$

$$u_d = \varepsilon u_{d1} + \varepsilon^2 u_{d2} + \varepsilon^3 u_{d3} + \cdots \tag{14}$$

$$\emptyset = \varepsilon \emptyset_1 - \varepsilon^2 \emptyset_2 + \varepsilon^3 \emptyset_3 - \cdots \tag{15}$$

$$p = 1 + \varepsilon p_{d1} + \varepsilon^2 p_{d2} + \varepsilon^3 p_{d3} + \cdots \tag{16}$$

$$z = 1 + \varepsilon^2 z_{d1} + \varepsilon^4 z_{d2} + \cdots \tag{17}$$

Substituting equation (13-17) into equations (1-4) and collecting terms with same powers of $\varepsilon$, from the cofficients of lowest order we have:

$$n_{d1} = \frac{1}{\lambda^2 - 3\sigma_d} \emptyset_1 \, ; \, u_{d1} = \frac{\lambda}{\lambda^2 - 3\sigma_d} \emptyset_1 \, ; \, p_{d1} = \frac{3}{\lambda^2 - 3\sigma_d} \emptyset_1 \, ; \, \eta = \frac{1}{3\sigma_d - \lambda^2} = \frac{\mu \sigma_i + 3\alpha\mu \sigma_i - \alpha + 1}{(1-\mu)(1+3\alpha)} \tag{18}$$

And from the higher order coefficients of $\varepsilon$ we have:

$$-\lambda \frac{\partial n_{d2}}{\partial \xi} + \frac{\partial n_{d1}}{\partial \tau} + \frac{\partial u_{d2}}{\partial \xi} + \frac{\partial (n_{d1} u_{d1})}{\partial \xi} = 0 \tag{19}$$

$$-\lambda \frac{\partial u_{d2}}{\partial \xi} - \lambda n_{d1} \frac{\partial u_{d1}}{\partial \xi} + \frac{\partial u_{d1}}{\partial \tau} + u_{d1} \frac{\partial u_{d1}}{\partial \xi} + \sigma_d \frac{\partial p_{d2}}{\partial \xi} = \frac{\partial \emptyset_2}{\partial \xi} - n_{d1} \frac{\partial \emptyset_1}{\partial \xi} \tag{20}$$

$$-\lambda \frac{\partial p_{d2}}{\partial \xi} + \frac{\partial p_{d1}}{\partial \tau} + u_{d1} \frac{\partial p_{d1}}{\partial \xi} + 3 \frac{\partial u_{d2}}{\partial \xi} + 3 p_{d1} \frac{\partial u_{d1}}{\partial \xi} = 0 \tag{21}$$

$$\frac{\partial^2 \emptyset}{\partial \xi^2} = z_{d1} + n_{d2} - \left(\frac{\mu\sigma_i + 3\alpha\mu\sigma_i - \alpha + 1}{(1-\mu)(1+3\alpha)}\right)\emptyset_2 + \emptyset_1^2 \left\{\frac{\mu\sigma_i^2 - 1}{2(1-\mu)}\right\} \tag{22}$$

By substituting equation (19-22) into (18) kdv equation is obtained:

$$\frac{\partial \emptyset_1}{\partial \tau} + C\emptyset_1 \frac{\partial \emptyset_1}{\partial \xi} + D\frac{\partial^3 \emptyset_1}{\partial \xi^3} = 0 \tag{23}$$

Where coefficient C can be positive or negative. With positive C, solitary wave is compressive and with negative C, solitary wave is rarefactive. The coefficients C and D are functions of σ, α and μ and given by:

$$C = \frac{-\lambda}{-\eta + 3\eta^2 \sigma_d + \eta^2 \lambda^2} \tag{24}$$

$$D = \frac{3\eta^2 \lambda + 2A\lambda - 12\eta^3 \sigma_d \lambda}{-\eta + 3\eta^2 \sigma_d + \eta^2 \lambda^2} \tag{25}$$

Where σ, α and μ are plasma parameters. It can be shown that equation (23) to possess one- or more localized soliton solutions(see e.g. in Refs. [15,16]). The simplest one soliton solution has the pulsed-shaped form

$$\emptyset_1 = \emptyset_m \text{sech}^2 \left(\frac{x}{\omega}\right) \tag{26}$$

$$\emptyset_m = \frac{3\lambda}{C} \; ; \; \omega = 2\sqrt{\frac{D}{\lambda}} \; ; \; \chi = \xi - \lambda\tau \tag{27}$$

Where $\emptyset_m = \frac{3\lambda}{C}$ is soliton amplitude, $\omega = 2\sqrt{\frac{D}{\lambda}}$ is soliton width and $\lambda$ is soliton velocity.

To consider solitonic solution related to a frame which moves with velocity , we define a variable $\chi = l\xi - \lambda\tau$, where $l$ is directional cosin of the wave vector $k$ along the $\xi$.

By integrating the new form of Eq. (23) respect to the variable χ and using the vanishing boundary condition for $\emptyset_1$ and its derivatives up to the second order for $|\chi| \to \infty$, we have[17]:

$$\frac{d^2 \emptyset_1}{d\chi^2} = \frac{\lambda}{Dl^3}\emptyset_1 - \frac{C}{2Dl^2}\emptyset_1^2 \tag{28}$$

This equation have solitonic solution as:

$$\emptyset_1 = \emptyset_m \text{sech}^2 \left(\frac{\chi}{\omega}\right) \tag{29}$$

$$\emptyset_m = \frac{3\lambda}{Cl} \; ; \; \omega = 2\sqrt{\frac{Dl^3}{\lambda}} \; ; \; \chi = \xi l - \lambda\tau \tag{30}$$

Where as the previous $\emptyset_m$ is soliton amplitude and $\omega$ is soliton width. As it can be seen, the soliton amplitude is inversely proportional to directional cosin but the soliton width is directly proportional to directional cosin. The width of a stable solitary wave solution must be real.

The changes of soliton shape is investigated for different values of $\alpha, \sigma_i, \mu, \lambda$ and $\sigma_d$, and plotted in figures 1-7. As can be seen in figure 1 the sign of amplitude is changed with increasing $\mu$, also the amplitude and width is increased and decreased respectively, with increasing $\mu$. Figures 2-5 also shows the same result. Figure 6 show that the amplitude and width are decreased and increased with increasing $\lambda$ respectively.

The stability conditions of this solution is analysed based on the simplified energetic approach[18]. The nature of solutions and the stability conditions may be treated by introducing Sagdeev potential. Equation (28) can be rewritten in the form:

$$\frac{d^2\emptyset_1}{d\chi^2} = \frac{\lambda}{Dl^3}\emptyset_1 - \frac{C}{2Dl^2}\emptyset_1^2 = -\frac{dV(\emptyset_1)}{d\emptyset_1} \tag{31}$$

equation (31) yields the nonlinear equation of motion as:

$$\frac{1}{2}\left(\frac{d\emptyset_1}{d\chi}\right)^2 + V(\emptyset_1) = 0 \tag{32}$$

and the Sagdeev potential $V(\emptyset_1)$ is obtained as:

$$V(\emptyset_1) = \frac{C}{6Dl^2}\emptyset_1^3 - \frac{\lambda}{Dl^3}\emptyset_1^2 \tag{33}$$

The stable solitonic solution of (31) exist if we have[19,20]:

1) $\left.\frac{d^2V(\emptyset_1)}{d\emptyset_1^2}\right|_{\emptyset_1=0} < 0$ (34)

2) A nanozero crossing point $\emptyset_1$ must exists so that

$$V(\emptyset_1 = \emptyset_0) = 0 \tag{35}$$

3) A point such $\emptyset_1$ must exists between 0 and $\emptyset_0$ so that $V(\emptyset_1) < 0$.

It can be seen at $\emptyset_1 = 0$ we have:

$$V(\emptyset_1) = \frac{dV(\emptyset_1)}{d\emptyset_1} = 0 \tag{36}$$

Therefore from equations (32) and (33) we have:

$$\left.\frac{d^2V(\emptyset_1)}{d\emptyset_1^2}\right|_{\emptyset_1=0} = -\frac{\lambda}{Dl^3} < 0 \tag{37}$$

By considering $D > 0$ (when $\lambda \to 0, \lambda > 0$ *or* $\sigma_d \to 0$), if $\theta$, the angle between wave vector $k$ and $\xi$ axis, varies from 0 to 90, therefore soliton is stable and as well as if $D < 0$ and also $\theta$, varies between 90 to 180, therefore soliton is stable.

## 4. The modified kdv equation:

The C coefficient of nonlinear term can be positive or negative. If we choose C=0 the amplitude of the wave becomes infinite therefore independent variables:

$$\xi = \varepsilon^{\frac{1}{2}}(x - \lambda t) \text{ and } \tau = \varepsilon^{\frac{3}{2}}t \tag{38}$$

is not valid and consequently we define new independent variables as

$$\xi = \varepsilon(x - \lambda t) \quad , \quad \tau = \varepsilon^3 t \tag{39}$$

substituting equation (13-17) into (1-4) and collecting terms with the same power of $\varepsilon$, from the coefficients of lowest order we have;

$$n_{d1} = \frac{1}{\lambda^2 - 3\sigma_d}\emptyset_1; \; u_{d1} = \frac{\lambda}{\lambda^2 - 3\sigma_d}\emptyset_1; \; p_{d1} = \frac{3}{\lambda^2 - 3\sigma_d}\emptyset_1; \; N = \frac{1}{3\sigma_d - \lambda^2} = \frac{\mu\sigma_i + 3\alpha\mu\sigma_i - \alpha + 1}{(1-\mu)(1+3\alpha)} \tag{40}$$

And from the coefficients of second power of $\varepsilon$ we obtain:

$$-\lambda \frac{\partial n_{d2}}{\partial \xi} + \frac{\partial u_{d2}}{\partial \xi} + \frac{\partial (n_{d1}u_{d1})}{\partial \xi} = 0 \tag{41}$$

$$-\lambda \frac{\partial u_{d2}}{\partial \xi} - \lambda n_{d1} \frac{\partial u_{d2}}{\partial \xi} + u_{d1} \frac{\partial u_{d1}}{\partial \xi} + \sigma_d \frac{\partial p_{d2}}{\partial \xi} = \frac{\partial \emptyset_2}{\partial \xi} - n_{d1} \frac{\partial \emptyset_1}{\partial \xi} \tag{42}$$

$$-\lambda \frac{\partial p_{d2}}{\partial \xi} + u_{d1} \frac{\partial p_{d1}}{\partial \xi} + 3 \frac{\partial u_{d2}}{\partial \xi} + 3p_{d1} \frac{\partial u_{d1}}{\partial \xi} = 0 \tag{43}$$

$$z_{d1} + n_{d2} - \left(\frac{\mu\sigma_i + 3\alpha\mu\sigma_i - \alpha + 1}{(1-\mu)(1+3\alpha)}\right)\emptyset_2 + \emptyset_1^2 \left\{\frac{\mu\sigma_i^2 - 1}{2(1-\mu)}\right\} = 0 \tag{44}$$

also the coefficients of higher power of $\varepsilon$ gives:

$$-\lambda \frac{\partial n_{d3}}{\partial \xi} + \frac{\partial n_{d1}}{\partial \tau} + \frac{\partial u_{d3}}{\partial \xi} + \frac{\partial (n_{d1}u_{d2})}{\partial \xi} + \frac{\partial (n_{d2}u_{d1})}{\partial \xi} = 0 \tag{45}$$

$$\frac{\partial u_{d1}}{\partial \tau} - \lambda n_{d2} \frac{\partial u_{d1}}{\partial \xi} - \lambda n_{d1} \frac{\partial u_{d2}}{\partial \xi} - \lambda \frac{\partial u_{d3}}{\partial \xi} + n_{d1}u_{d1} \frac{\partial u_{d1}}{\partial \xi} + u_{d2} \frac{\partial u_{d1}}{\partial \xi} + u_{d1} \frac{\partial u_{d2}}{\partial \xi} + \sigma_d \frac{\partial p_{d3}}{\partial \xi} = -\frac{\partial \emptyset_3}{\partial \xi} - z_{d1} \frac{\partial \emptyset_1}{\partial \xi} + n_{d1} \frac{\partial \emptyset_2}{\partial \xi} - n_{d2} \frac{\partial \emptyset_1}{\partial \xi} \tag{46}$$

$$-\lambda \frac{\partial p_{d3}}{\partial \xi} + \frac{\partial p_{d1}}{\partial \tau} + u_{d1} \frac{\partial p_{d2}}{\partial \xi} + u_{d2} \frac{\partial p_{d1}}{\partial \xi} + 3 \frac{\partial u_{d3}}{\partial \xi} + 3p_{d1} \frac{\partial u_{d2}}{\partial \xi} + 3p_{d2} \frac{\partial u_{d1}}{\partial \xi} = 0 \tag{47}$$

$$\frac{\partial^2 \emptyset_1}{\partial \xi^2} = n_{d3} + n_{d1}z_{d1} + \left(\frac{\mu\sigma_i + 3\alpha\mu\sigma_i - \alpha + 1}{(1-\mu)(1+3\alpha)}\right)\emptyset_3 + \left\{\frac{1-\mu\sigma_i^2}{(1-\mu)}\right\}\emptyset_1\emptyset_2 - \frac{2\alpha}{(1-\mu)(1+3\alpha)}\emptyset_1^3 \tag{48}$$

By introducing:

$$A = -\frac{1}{2}\left\{\frac{1-\mu\sigma_i^2}{(1-\mu)}\right\} \quad and \quad B = -\frac{2\alpha}{(1-\mu)(1+3\alpha)} \tag{49}$$

and by substituting equations (40) and (41-44) into (45-48) the following equation is obtained:

$$\frac{\partial \emptyset_1}{\partial \tau} + F \frac{\partial \emptyset_1}{\partial \xi} + G \frac{\partial (\emptyset_1\emptyset_2)}{\partial \xi} + H\emptyset_1^2 \frac{\partial \emptyset_1}{\partial \xi} + K \frac{\partial^3 \emptyset_1}{\partial \xi^3} = 0 \tag{50}$$

Where

$$F = \frac{12N^2\sigma_d\lambda z_{d1} - N\lambda z_{d1}}{N - 3N^2\sigma_d - N^2\lambda^2}; K = \frac{\lambda}{N - 3N^2\sigma_d - N^2\lambda^2}; G = \frac{3N^2\lambda + 2A\lambda - 12N^3\sigma_d\lambda}{N - 3N^2\sigma_d - N^2\lambda^2} \tag{51}$$

$$H = \frac{-7N\lambda A + 36N^2\lambda A\sigma_d - 3B\lambda - 2N^2\lambda^3 A - 2N^4\lambda^3 + 2N^2\lambda^2 A + 2N^4\lambda^2 - 12N^4\lambda\sigma_d}{N - 3N^2\sigma_d - N^2\lambda^2} \tag{52}$$

It is clear that G = C , so far critical parameters "G " becomes zero and in this situation Eq. (50) reduces into the mkdv equation:

$$\frac{\partial \emptyset_1}{\partial \tau} + F\frac{\partial \emptyset_1}{\partial \xi} + H\emptyset_1^2 \frac{\partial \emptyset_1}{\partial \xi} + K\frac{\partial^3 \emptyset_1}{\partial \xi^3} = 0 \tag{53}$$

The solitonic solution of equation (53) is as follows:

$$\emptyset_1 = \pm\emptyset_m \operatorname{sech}\left(\frac{\chi}{\omega}\right) \tag{54}$$

$$\chi = \xi - \lambda\tau \quad ; \quad \emptyset_m = \sqrt{\frac{6(\lambda-F)}{H}} \quad ; \quad \omega = \sqrt{\frac{K}{\lambda-F}} \tag{55}$$

Where $\emptyset_m$ is soliton amplitude and $\omega$ is soliton width.

## 5. Conclusion

We consider small amplitude dust acoustic solitary waves in dusty plasma. We obtained kdv and mkdv equations and also Sagdeev potential in unmagnetized plasma with Boltzmann distributed electrons and non-thermal distributed ions considering variable dust charge and entering influence component of pressure . The sign of C coefficient can be positive or negative. If we suppose C coefficient equal to zero and by considering variables $\xi$ and $\tau$, mkdv equation, which have solitonic solutions may be derived. The variation of dust charge changes the nonlinear and dispersive terms, previous articles about kdv equation have not considered the variation of dust charge and pressure component. As well as the definition of independent variables are different. The effects of non-thermal ions, density and temperature on the behavior of the solitons are investigated.

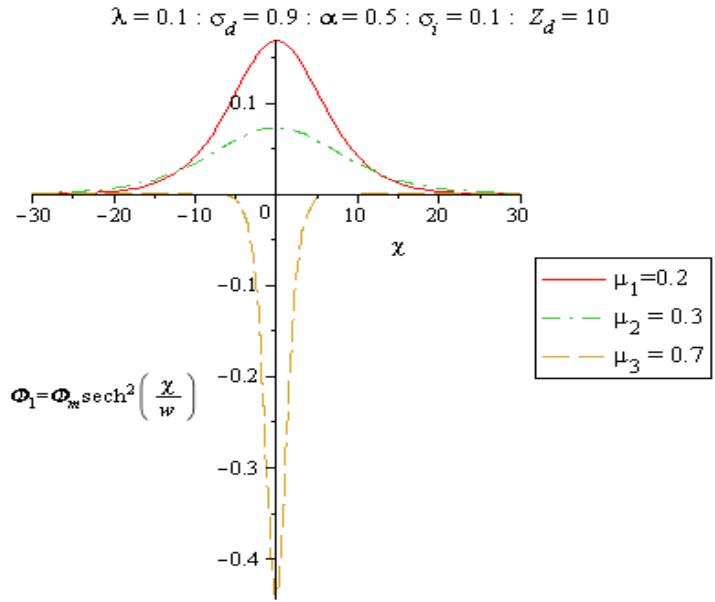

Fig. 1. $\Phi$ vs. x for fixed values of $\alpha, \sigma_i, \sigma_d, \lambda, z_d$ and different values of $\mu$.

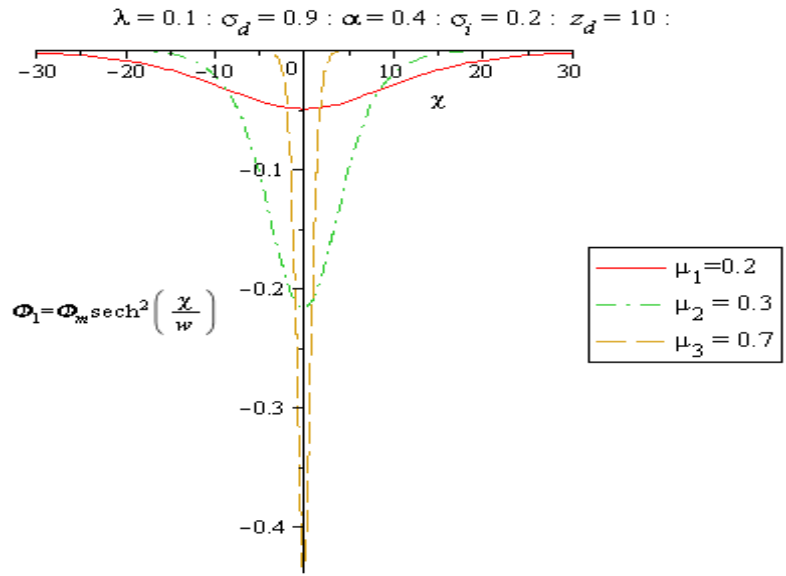

Fig. 2. Φ vs. x for fixed values of $\alpha, \sigma_i, \sigma_d, \lambda, z_d$ and different values of $\mu$.

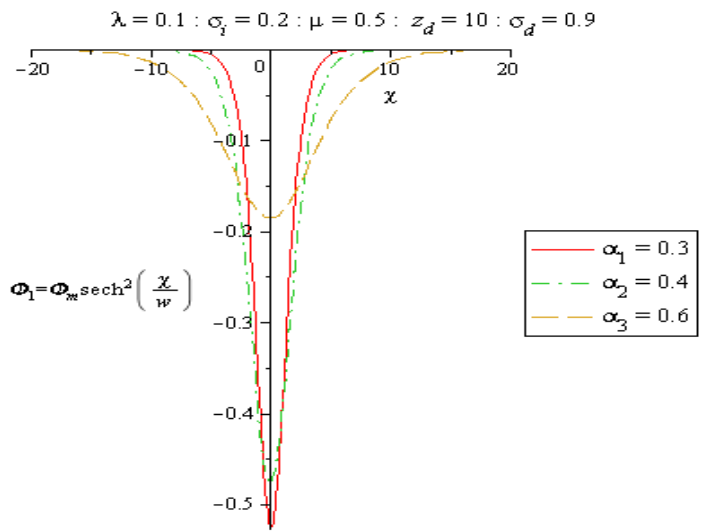

Fig. 3. Φ vs. x for fixed values of $\sigma_i, \sigma_d, \mu, z_d, \lambda$ and different values of $\alpha$.

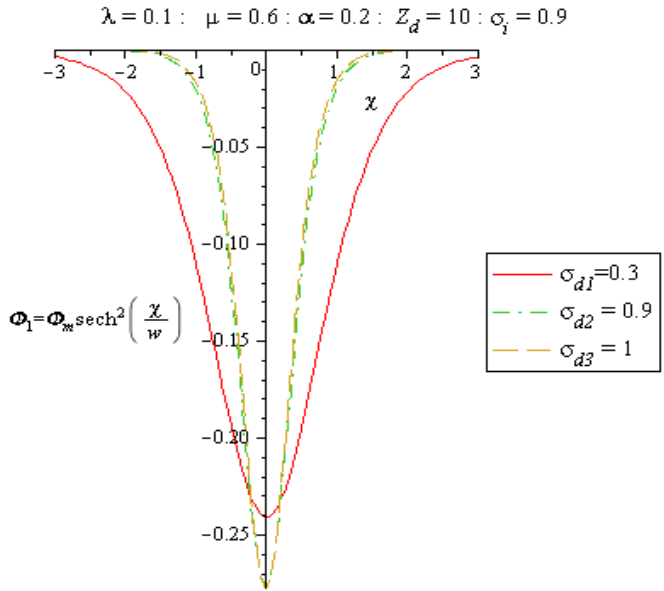

Fig. 4. Φ vs. x for fixed values of $\sigma_i, \alpha, \mu, z_d, \lambda$ and different values of $\sigma_d$.

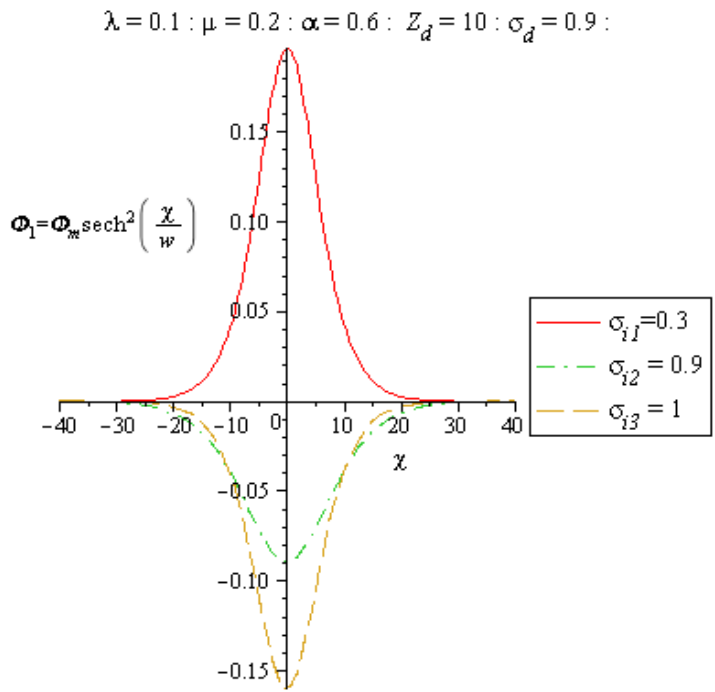

Fig. 5. Φ vs. x for fixed values of $\sigma_d, \lambda, \alpha, \mu, z_d,$ and different values of $\sigma_i$.

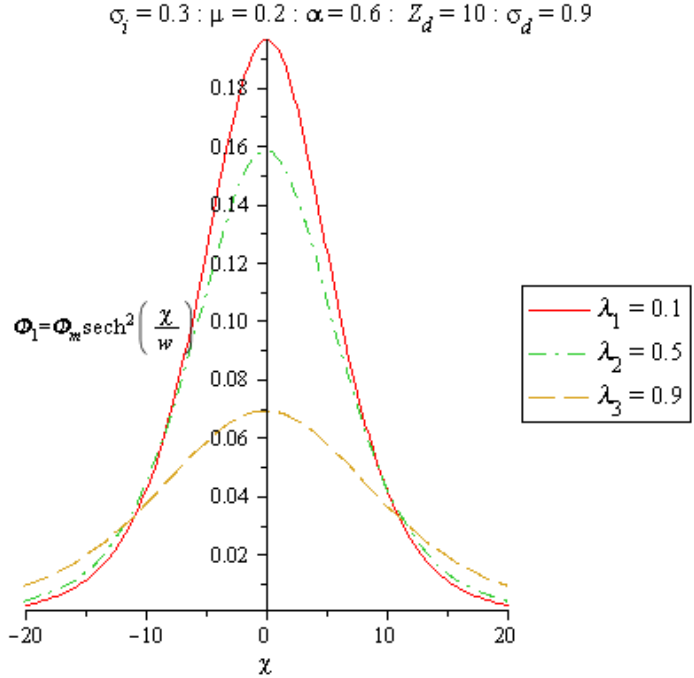

Fig. 6. $\Phi$ vs. x for fixed values of $\sigma_i, \sigma_d, \alpha, \mu, z_d,$ and different values of $\lambda$.